\documentclass[fleqn,twoside]{article}
\usepackage{espcrc2}

\usepackage{graphicx}
\usepackage[figuresright]{rotating}

\newcommand{\AmS}{{\protect\the\textfont2
  A\kern-.1667em\lower.5ex\hbox{M}\kern-.125emS}}

\title{Measurement of the Muon $(g-2)$-Value}

\author{B. Lee Roberts on behalf of the Muon $(g-2)$ 
                  Collaboration
\address{Department of Physics, Boston University,\\
        590 Commonwealth Avenue, Boston, MA 02215 USA }
                \thanks{R.M. Carey, 
E. Efstathiadis, M.F. Hare, X. Huang, F. Krienen,
A. Lam, I. Logashenko, 
J.P. Miller, J. Paley, Q. Peng, 
O. Rind, B.L. Roberts, L.R. Sulak, A. Trofimov -
Boston University;
G. Bennett, H.N. Brown, G. Bunce,
G.T. Danby, R. Larsen, Y.Y. Lee,
W. Meng, J. Mi, W.M. Morse,
D. Nikas, C. \"Ozben,  R. Prigl,
Y.K. Semertzidis, D. Warburton - Brookhaven National Laboratory;
Y. Orlov - Cornell University;
K. Jungmann -KVI Groningen;
A. Grossmann, P. von Walter, G. zu Putlitz - University Heidelberg;
P.T. Debevec, W. Denninger, F. Gray
D.W. Hertzog, C. Onderwater, C. Polly, M. Sossong, D. Urner,
- University of Illinois, Urbana Champaign;
A. Yamamoto - KEK
B. Bousquet, P. Cushman, L. Duong, S. Giron, 
J. Kindem, I. Kronkvist, R. McNabb,
T. Qian, P. Shagin
- University of Minnesota;
V.P. Druzhinin, G.V. Fedotovich, D. Grigoriev,
B.I. Khazin, N.M. Ryskulov, 
Yu.M. Shatunov, E. Solodov - Budker Institute;
M. Iwasaki - Tokyo Institute of Technology;
H. Deng,  M. Deile, S.K. Dhawan, F.J.M. Farley, M. Grosse-Perdekamp,
V.W. Hughes, D. Kawall, J. Pretz, 
S.I. Redin, E. Sichtermann, A. Steinmetz - Yale University}
                \thanks{Supported in part by the USNSF and USDOE}}
    
\begin{document}

\begin{abstract}
The muon $(g-2)$ experiment is described, and the recent results are
presented. These results represent the final measurement for the
positive muon.
\vspace{1pc}
\end{abstract}

\maketitle

\section{Introduction}

The measurement of magnetic moments has been
important in advancing our knowledge
of sub-atomic physics since the famous 1921 paper of Stern,\cite{stern}
 which laid out the principles of what we now call the ``Stern-Gerlach
experiment''.  The experimental and theoretical developments in the
study of the electron's anomalous
magnetic moment represent one of the great success
stories of modern physics.   The experimental accuracy has
reached a relative accuracy
of $\sim 4$ parts in $10^9$ (parts per billion),\cite{vd} and the theory is
constrained by our knowledge of the fine-structure constant $\alpha$, rather
than by the 
eight-order and tenth-order QED calculations.\cite{kinalpha} 

The gyromagnetic ratio $g$ is defined by 
\begin{equation}
\vec \mu_s = g ( {e \over 2m} ) \vec s, 
\end{equation}
where $\vec s$ is the spin angular momentum, and $\vec \mu$ is the
magnetic moment resulting from this angular momentum.
The Dirac equation for point particles 
predicts that $g\equiv 2$, but radiative corrections
increase the value at the part per mil level.
The  magnetic moment $\mu$ is defined as
as $\mu = (1+a) {e \hbar / 2 m}$, where  $a = (g - 2)/ 2$
is the anomalous magnetic moment (or simply the anomaly).

When E821 began in the early 1980s, 
$a_{\mu}$ was known to 7.3 parts per million (ppm).\cite{cern3}  
The E821 Collaboration
has reported four new measurements with relative accuracies of 13, 5,
1.3 and 0.7 ppm respectively.\cite{carey,brown1,brown2,bennett}

To the level of the experimental accuracy, the electron anomaly can be
described by the QED of $e^{\pm}$ and photons, with the contribution of
heavier virtual particles entering at a level of around
3 ppb.  The larger mass of
the muon permits heavier virtual particles to contribute, and the enhancement
factor is $ {\sim  ( {m_{\mu} /  m_e} )^2} \sim 40,000$.  The CERN
measurement observed the effect on $a_{\mu}$  of virtual hadrons 
at the 8 standard deviation level.\cite{cern3}
The standard model value of $a_{\mu}$ consists of 
QED, strong interaction and  weak radiative corrections,
and a significant 
deviation from the calculated standard model value would represent
a signal for non-standard model physics.  At this conference we
have heard much discussion of the theory of $(g-2)$,\cite{cz,se,md,ah}
so it will not be discussed further in this talk.\cite{dehz}

\section{The Experimental Technique}

The method used in the third CERN experiment and the BNL experiment are
very similar, save the use of direct muon injection into the storage ring
which was developed by the E821 collaboration.  These
experiments are based on the
fact that for $g\neq 2$ (or more precisely $a_{\mu} > 0$) the spin 
precesses faster than
the momentum vector when a muon travels transversely to a 
magnetic field.  The Larmor and Thomas spin-precession and the momentum 
precession frequencies are
\begin{equation}
 \omega_S = {geB \over 2 m c} + (1-\gamma) {e B \over \gamma mc};\qquad
 \omega_C = {e B \over mc \gamma}
\end{equation}
and the difference frequency gives the frequency with which the spin
precesses relative to the momentum, 
\begin{equation}
 \qquad
\omega_a = \omega_S - \omega_C = ({g-2 \over 2}) {eB \over mc}
\label{eq:omeganoE}
\end{equation}
which is proportional to the anomaly, rather than to the full magnetic
moment. A precision measurement of $a_{\mu}$ requires precision measurements
of the precession frequency $\omega_a$  and the magnetic field,
which is expressed as the free-proton precession frequency
$\omega_p$ in the storage ring magnetic field.

The muon frequency can be measured as accurately as the counting
statistics and detector apparatus permit.  
The design goal for the NMR magnetometer and calibration system
was a field accuracy of about 0.1 ppm.  The $B$ which enters in 
Eq. \ref{eq:omeganoE} is the average field seen by the ensemble of muons
in the storage ring, 
$<B>_{\phi} 
=$  $<{1\over \pi R^2} \int_0^R \int_0^{2\pi} M(r,\theta)B(r, \theta) rdr
d\theta>_{\phi}$ 
where $\phi$ is the azimuthal angle around the ring, $r,\theta$ are the
coordinates at a single slice of azimuth centered at the middle of the
90 mm diameter muon storage region.  $M(r,\theta)$ is the moment (multipole)
distribution of the muon beam, and couples multipole by multipole
with the magnetic field multipoles.  In the analysis, the field
is averaged over the data collection time as well.

The need for vertical focusing implies that a gradient field is needed,
but the usual magnetic gradient used in storage rings is ruled out in
our case.  A sufficient magnetic gradient for vertical focusing would
spoil the ability to use NMR to 
measure the magnetic field to the necessary accuracy. Furthermore,
it is very difficult to obtain adequate information
on the higher
moments of the muon distribution in the storage ring, so the presence
of higher multipoles in $<B>$ is also undesirable for this
reason. A round beam-profile was chosen, since sharp corners
would imply large higher moments for $M(r,\theta)$. 

An electric quadrupole is used for vertical focusing, 
taking advantage of the 
``magic''~$\gamma=29.3$ at which an electric field does not contribute to
the spin motion relative to the momentum. In the presence of an electric
and a magnetic field, the spin difference frequency is given by
\begin{equation}
\vec \omega_a = {e \over mc}
\left[ a_{\mu} \vec B -
\left( a_{\mu}- {1 \over \gamma^2 - 1}\right) \vec \beta \times \vec E
\right],
\label{eq:tbmt}
\end{equation}
which reduces to Eq. \ref{eq:omeganoE} in the absence of an electric field.
Note that for muons with $\gamma = 29.3$ in an electric field alone,
the spin would follow the momentum vector.

The arrangement of a magnetic dipole field combined with an electric
quadrupole field is called a Penning trap in atomic physics.
However with a 14 m diameter and $\sim700$ T weight, 
the scale of our trap is
quite different from the usual one.\cite{vd}  

In order to meet the conditions discussed above, a goal of $\pm 1$ ppm
uniformity of the \break $<B>$-field over the storage region was set and met. 
Given the projected knowledge of the
muon distribution, the allowable strength of the 
quadrupole and higher magnetic
multipoles was also determined. 

A kick of about 0.1 Tm is needed to bring the beam onto a stable orbit.
This is achieved with three 1.7 m long ferrite-free kickers,\cite{kicker} 
which can be
thought of as single-loop pulsed magnets carrying a current of 
4,200 A.  The minimum inductance achievable of 1.6 $\mu$H 
limited the peak current to 4200A, and resulted in a current-pulse 
width $\sim2.5$ times greater
than the cyclotron period of 150 ns.  The phase space mis-match between the 
size of the inflector exit and the storage region, and
multiple scattering in the inflector end, reduces the calculated
injection efficiency to $\sim 8.7$\%.
The less than optimal kicker pulse further reduces the
injection efficiency to about 7.3\%.  Nevertheless the effective data
rate
per fill is almost a factor of 100 over that available in the final CERN
experiment,\cite{cern3} a factor of 10 coming from muon injection, and
a factor of 10 coming from the AGS intensity.  With direct muon injection,
the injection-related background seen by the
detectors is down by a factor of 50.\cite{carey,brown1}

The $(g-2)$ ring functions as a weak focusing storage ring with the
field index 
\begin{equation}
n = {\kappa R_0 \over \beta B_0},
\label{eq:n}
\end{equation}
where $ \kappa$ is the electric quadrupole  gradient.  Several $n$~-~values
were used for data acquisition: 
$n = 0.137,\ 0.142$ and 0.122, the latter two having been used for $\mu^-$. 
The horizontal (radial) and 
vertical betatron frequencies are (approximately) given by
\begin{equation}
 f_x = f_C \sqrt{1-n}\simeq 0.93 f_C;\ 
f_y = f_C \sqrt{n} \simeq 0.37 f_C
\label{eq:betafreq}
\end{equation}
where $f_C$ is the cyclotron frequency and the numerical values 
assume $n=0.137$.

The experimental signal is the $e^{\pm}$ from $\mu^{\pm}$ decay, which 
were detected by lead-scintillating
fiber calorimeters.\cite{det}  The time and energy of each event was
 stored for analysis offline. 
Muon decay is a three-body decay, so the 3.1 GeV muons produce a continuum
of positrons (electrons) from the end-point energy down.  Since the highest
energy  $e^{\pm}$ are correlated with the muon spin, if one counts high energy 
 $e^{\pm}$ as a function of time, one gets an exponential from muon decay
modulated by the $(g-2)$ precession. The expected form for the positron time
spectrum is 
\begin{equation}
f(t) =  {N_0} e^{- \lambda t } 
[ 1 + {A} \cos ({\omega_a} t + {\phi})] 
\label{eq:5pm}
\end{equation}

\begin{figure}[htb]
\centering
  \includegraphics*[width=54mm]{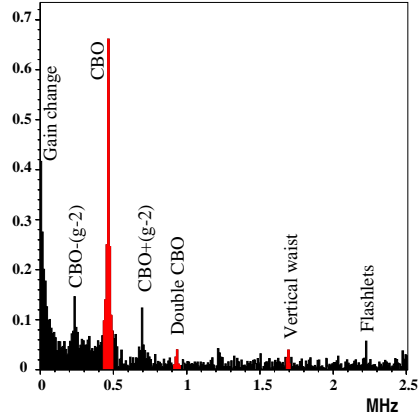}
  \caption{A Fourier Transform of the fit residuals from a 5-parameter fit to
the 1999 data set.  To adequately describe the data, additional parameters
were necessary (see Eq. \ref{eq:robjim}).
       \label{fg:FT5pm}}
\end{figure}

However, a Fourier analysis of the residuals from this five parameter 
fit (see  Fig. \ref{fg:FT5pm})
shows a number of frequency components which can be understood
from the beam dynamics in the ring.\cite{bennett,epac}  
The most prominent frequencies 
in the residuals come from the coherent oscillation of the beam.

\begin{figure}[!ht]
\centering
  \includegraphics*[width=4.4cm]{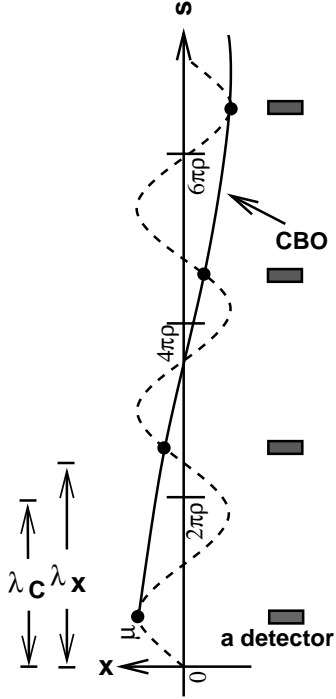}
  \caption{A cartoon showing the beam motion (dashed line), 
and the apparent motion (solid line)
seen by a single detector.
The vertical axis is distance along the 
storage ring (shown for 4 turns) and the horizontal axis shows the 
radial betatron amplitude.  $\lambda_C = 2 \pi \rho$ 
is the cyclotron wavelength,
$\lambda_x$ is the radial betatron wavelength, and $\rho$ is the
central orbit radius.
       \label{fg:cbocartoon}}
\end{figure}

While the frequencies given in Eq. \ref{eq:betafreq} describe
the motion of a single beam particle, the aperture mis-match between
the inflector exit and the storage aperture, along with an imperfect
kick and momentum dispersion in the ring,
will cause 
the beam to undergo coherent radial oscillations. 
The detector acceptance depends on the radial position of the muon
when it decays, so any coherent radial beam
motion will amplitude modulate the decay $e^{\pm}$ distribution.
The principal frequency will be the
 ``Coherent Betatron Frequency''
\begin{equation}
f_{\rm CBO} = f_C - f_x = (1 - \sqrt{1-n})f_C 
\label{eq:cbo}
\end{equation}
which is the frequency a single fixed detector sees the beam moving
 coherently back and forth. This motion can be understood by the 
cartoon in Fig. \ref{fg:cbocartoon}.
 It is this CBO frequency and its
sidebands from beating with the $(g-2)$ frequency which 
are clearly visible in the Fourier spectrum of Fig.~\ref{fg:FT5pm}.
$n = 0.137$ was chosen to avoid storage ring resonances, however
the resulting CBO frequency was close to the second harmonic of 
$f_a(=\omega_a/2\pi)$ putting the difference sideband close to $f_a$.
The $n$-value was changed for the 2001 data collection period to reduce 
our sensitivity to this difficulty.

The CBO modified the positron time spectrum by
\begin{equation}
N_p = N_0 e^{-t\over \tau}(1+A'\sin{(\omega_a t + \phi')}) \ \ 
\times \label{eq:robjim}
\end{equation}
$$
(1+A_{CBO}(t)\cos{(\omega_{CBO}t+\phi_{CBO})}),
$$
 where
\begin{equation}
A'=A(1+A_1(t)\cos{(\omega_{CBO}t+\phi_1)})
\end{equation}
\begin{equation}
\phi'=\phi(1+A_2(t)\cos{(\omega_{CBO}t+\phi_2)})
\end{equation}
The terms
$A_1$ and $A_2$ can cause artificial shifts in $\omega_a$
up to 4 ppm in individual detectors when not accounted for in
the fitting procedure.  The final fitting proceedure included these,
and other effects such as pulse pile-up, detector gain changes and
muon losses from the ring other than by decay.

\section{Results and Conclusions}

\begin{figure}[!ht]
\centering
\includegraphics*[width=3.4in]{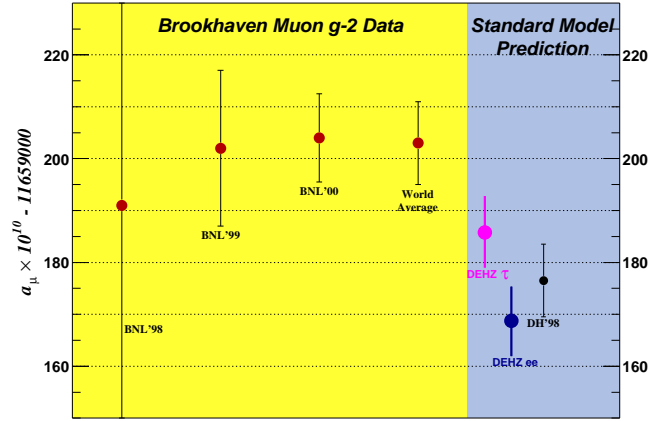}
  \caption{E821 measurements of $a_{\mu}$ carried out with direct muon
injection into the storage ring.  The relative uncertainties are
$\pm 5$ ppm (98), $\pm 1.3$ ppm (99), $\pm 0.7$ ppm (00), where the 
number in parentheses is the year when the data were collected.  All
measurements are for $\mu^+$. The three theory points use the
lowest order hadronic contribution from DH98\cite{dh98} and the separate
DEHZ evaluations\cite{dehz} from
$\tau$ and $e^+e^-$  hadronic data.  The DH98 evaluation
includes both $\tau$ and 
$e^+e^-$ data.  
       \label{fg:results}}
\end{figure}

After consistent results were obtained
in four independent (and three different)
analysis procedures for $\omega_a$, and two independent studies of
$\omega_p$, the offsets were removed and the value 
\begin{equation}
  a_{\mu^+} =  11\,659\,204(7)(5)\ \times\ 10^{-10}~~\mbox{(0.7\,ppm)}
\end{equation}
was obtained.\cite{bennett}  Excellent agreement was found with the
previous measurements from Brookhaven and from CERN.

The E821 results 
are displayed graphically in Fig. \ref{fg:results} along with the 
Davier-H\"ocker(98)\cite{dh98} first order 
hadronic contribution, and the most
recent results from $\tau$-decay and $e^+e^-$ annihilation reported at
this conference.\cite{se,md,ah,dehz}

Now that the $\tau$-decay value of $a_{\mu}{\rm (Had;1)}$ disagrees with
the value obtained from electron positron annihilation,
the situation is somewhat confused.  While the path from hadronic
electron-positron annihilation to a value of $a_{\mu}{\rm Had;1}$
is somewhat more theoretically direct than from hadronic tau decay,
the extraction of $a_{\mu}{\rm( Had;1)}$ from
tau-decay data has been carefully studied, with all the expected 
effects included.  

  However, as we have heard at this meeting,
there are systematic differences between other quantities
when $e^+e^-$ annihilation and $\tau$-decay are compared.
Since the low-energy $e^+e^-$ data are dominated by the
recent precise data from Novosibirsk, it is fortunate
that further work on
$e^+e^-$-annihilation is being carried out at Frascati, Belle and
BaBar, and 
they will either confirm or disagree with the recent Novosibirsk
data.  Comparison with the new DEHZ evaluations\cite{dehz} shows either
a 1.6  or 3 standard deviation discrepancy. An independent analysis\cite{hag}
of the  $e^+e^-$ data agrees well with the equivalent 
DEHZ analysis.  Thus the significance
for an indication of new physics will have to wait for clarification
of the correctness of the hadronic contribution.

Nevertheless, a recent very conservative evaluation of the impact
of $(g-2)$ on the constraining of supersymmetry parameters shows
that even with the current uncertainties, $(g-2)$ already rules out
 a ``substantial region of (susy) parameter space...
that has not been probed by any previous experiment''.\cite{mw}

The E821 collaboration has one additional data set, which was taken
with $\mu^-$ in the ring.  The experimental uncertainty will be
on the order of 0.8 ppm.  
Obviously we will have to stay tuned for further clarification on
the hadronic contribution, and the analysis of the  $\mu^-$ data.

\end{document}